\documentclass[]{aastex63}

\submitjournal{AJ}

\shorttitle{Prediction model of soft proton intensities}
\shortauthors{Kronberg et al.}
\usepackage{amsmath,amssymb}
\usepackage{booktabs}

\include{notation}

\begin{document}

\title{Prediction of soft proton intensities in the near-Earth space using machine learning }

\correspondingauthor{Elena Kronberg}
\email{elena.kronberg@lmu.de}

\author[0000-0001-7741-682X]{Elena A. Kronberg}
\affiliation{Department of Earth and Environmental Sciences(Geophysics), Ludwig Maximilian University of Munich (LMU) Munich,
Theresienstr. 41,
Munich, 80333, Germany}

\author[0000-0001-9957-9531]{Tanveer Hannan}
\affiliation{Institute of Informatics, LMU Munich, Oettingenstraße 67, Munich, D-80538, Germany}
\affiliation{Department of Statistics, LMU Munich, Ludwigstraße 33, Munich, D-80539, Germany}

\author[0000-0001-7223-4103]{Jens Huthmacher}
\affiliation{Institute of Informatics, LMU Munich, Oettingenstraße 67, Munich, D-80538, Germany}

\author[0000-0002-8712-4035]{Marcus Münzer}
\affiliation{Institute of Informatics, LMU Munich, Oettingenstraße 67, Munich, D-80538, Germany}

\author[0000-0002-7070-1078]{Florian Peste}
\affiliation{Institute of Informatics, LMU Munich, Oettingenstraße 67, Munich, D-80538, Germany}

\author[0000-0003-0154-6948]{Ziyang Zhou}
\affiliation{Institute of Informatics, LMU Munich, Oettingenstraße 67, Munich, D-80538, Germany}

\author[0000-0001-9724-4009]{Max Berrendorf}
\affiliation{Institute of Informatics, LMU Munich, Oettingenstraße 67, Munich, D-80538, Germany}
	
	\author[0000-0001-8841-5128]{Evgeniy Faerman}
\affiliation{Institute of Informatics, LMU Munich, Oettingenstraße 67, Munich, D-80538, Germany}

%\author{al.}
\author[0000-0002-9112-0184]{Fabio Gastaldello}
\affiliation{Instituto di Astrofisica Spaziale e Fisica Cosmica (INAF-IASF), Milano, via A. Corti 12, I-20133 Milano, Italy}

\author[0000-0003-0879-7328]{Simona Ghizzardi}
\affiliation{Instituto di Astrofisica Spaziale e Fisica Cosmica (INAF-IASF), Milano, via A. Corti 12, I-20133 Milano, Italy}

\author[0000-0003-4475-6769]{Philippe Escoubet}
\affiliation{European Space Research and Technology Centre, Noordwjik,  Keplerlaan 1, 2201 AZ, Netherlands}

\author[0000-0002-1241-7570]{Stein Haaland}
\affiliation{Birkeland Centre for Space Science, University of Bergen, Allégaten 55, NO-5007 Bergen, Norway}
\affiliation{Max Planck Institute for Solar System Research,
Justus-von-Liebig-Weg 3, Göttingen, 37077, Germany}

\author[0000-0003-3689-4336]{Artem Smirnov}
\affiliation{German Research Centre for Geosciences, Albert-Einstein-Straße 42-46, Potsdam, D-14473, Germany}

\author[0000-0003-4278-0482]{Nithin Sivadas}
\affiliation{Department of Electrical and Computer Engineering, Boston University, 8 Saint Mary’s Street, Boston, MA 02134, USA}

\author[0000-0003-2079-5683]{Robert C. Allen}
\affiliation{Johns Hopkins University Applied Physics Lab, 11100 Johns Hopkins Road, Laurel, MD 20723, USA}

\author[0000-0002-6038-1090]{Andrea Tiengo}
\affiliation{Scuola Universitaria Superiore IUSS Pavia, piazza della Vittoria 15, I-27100 Pavia, Italy}
\affiliation{Instituto di Astrofisica Spaziale e Fisica Cosmica (INAF-IASF), Milano, via A. Corti 12, I-20133 Milano, Italy}
\affiliation{INFN, Sezione di Pavia, via A. Bassi 6, I-27100 Pavia, Italy}

\author[0000-0002-7305-2579]{Raluca Ilie}
\affiliation{University of Illinois at Urbana-Champaign, 306 N. Wright Street, 5054 ECEB,
	Urbana, IL 61801, USA}

\begin{abstract}
The spatial distribution of energetic protons contributes towards the understanding of magnetospheric dynamics. 
Based upon 17 years of the Cluster/RAPID observations, we have derived machine learning-based models to predict the proton intensities at energies from 28 to 1,885 keV in the 3D terrestrial magnetosphere at radial distances between 6 and 22 R$_\mathrm{E}$.
We used the satellite location and indices for solar, solar wind and geomagnetic activity as predictors. 
The results demonstrate that the neural network  (multi-layer perceptron regressor) outperforms baseline models based on the k-Nearest Neighbors and historical binning on average by $\sim$80\% and $\sim$33\%, respectively. 
The average correlation between the observed and predicted data is about 56\%, which is reasonable in light of the complex dynamics of fast-moving energetic protons in the magnetosphere. 
In addition to a quantitative analysis of the prediction results, we also investigate parameter importance in our model.
The most decisive parameters for predicting proton intensities are related to the location: ZGSE direction and the radial distance.
Among the activity indices, the solar wind dynamic pressure is the most important. 
The results have a direct practical application, for instance, for assessing the contamination particle background in the X-Ray telescopes for X-ray astronomy orbiting above the radiation belts.
To foster reproducible research and to enable the community to build upon our work we publish our complete code, the data, as well as weights of trained models.
Further description can be found in the GitHub project at \url{https://github.com/Tanveer81/deep_horizon}.

\end{abstract}


\section{Introduction}\label{sec:intro}
Understanding the distribution and dynamics of the energetic protons in the near-Earth space is essential not only for magnetospheric physics. Energetic protons are also suspected to damage space-based instrumentation and to affect their scientific performance. 
For example, X-ray telescopes such as Chandra~\citep{Weisskopf02} and X-ray Multi-Mirror Mission (XMM-Newton)~\citep{Jansen01} are suffering from contamination by so-called \emph{soft protons} (SP)~\citep{Deluca04,Leccardi08,Kuntz08}. 
These are protons at energies in the range of tens of keV up to a few MeV. The SPs that populate the solar wind (SW) and the Earth's magnetosphere can damage CCD detectors leading to a loss of available exposure time due to an increased background rate. 

Consequently, the performance of future X-Ray missions orbiting in the magnetosphere and the SW depends on how well the instruments are protected from the SP. 
For example, the Advanced Telescope for High-ENergy Astrophysics (ATHENA) mission~\citep{Nandra.ea:13} plans to deploy an array of magnets to deflect charged particles away from the instruments~\citep{Lotti.ea:18,Fioretti.ea:18}. 
Moreover, the original orbit choice might be changed from L2 to L1 due to a better understanding of this region's energetic particles' dynamics.
The Solar wind Magnetosphere Ionosphere Link Explorer (SMILE)~\citep{Raab16} mission also concerns the energetic particle levels in the magnetosheath that it will experience during its polar orbit.

There are few studies related to the energetic proton population in near-Earth space. 
In contrast to this work, they focus on well-confined regions around the Earth. 
For example,~\citet{meng81a,Kron12,Kron15} have studied the dependence of the distribution of energetic protons on SW and geomagnetic activity parameters in the plasma sheet. The plasma sheet is one of the few regions where it is expected to experience an enhanced background, also for XMM, see, for example,~\citet{Rosenqvist02}.
In~\citet{Kron15} the distribution is projected to the equatorial plane. 
However, the missions mentioned above plan on taking observations at high latitudes. 

Energetic protons can also be observed in other regions. 
The region upstream of the bow shock, for instance, is known to be populated with energetic protons due to the effective acceleration at the bow shock \citep[e.g.,][]{Lee82,Kron09}. 
SW phenomena such as Coronal Mass Ejections and Corotational Interaction Regions are associated with energetic charged particles observed by spacecraft in the region upstream of the bow shock. 
The bow shock also serves to heat the SW plasma as it flows into the magnetosheath, the boundary region between the bow shock and the magnetopause, resulting in harder energy spectrum.
This region is generally considered void of energetic particles ($\sim$100 keV).
However, occasionally phenomena associated with energetic particles such as hot flow anomalies, plasma jets~\citep[e.g.,][]{Facsko10,Savin14} occur.
Some of these energetic particles escape the magnetosphere~\citep[e.g.,][]{Kron11}. 
The diamagnetic cavities at cusps are also effective particle accelerators~\citep[e.g.,][]{Nykyri12}.

Studying each region in isolation enhances our knowledge about the physics in these regions and simplifies modeling. 
However, the boundaries between the regions are not always well defined, for instance, due to the quasi-parallel bow shock formation or Kelvin-Helmholtz instability.
Therefore, it makes sense for space weather applications to study the proton intensities in the near-Earth region holistically. 
Moreover, most previous studies only consider a few input solar or geomagnetic parameters instead of utilizing the full range of them.

\citet{Kron20a}~studied the dependence of SP contamination in the XMM-Newton telescope on location and various solar and geomagnetic parameters using a machine learning approach. 
The study revealed the strongest dependence of the contamination on the location and the SW velocity. Simultaneously, parameters such as the AE or SYM-H indices (which measure the geomagnetic activity, namely, the magnetic field disturbance in the auroral region of the northern hemisphere and the disturbance of the geomagnetic field at the equatorial regions, respectively)  have shown significantly lower importance levels for the contamination, which was rather unexpected from common knowledge of magnetospheric dynamics.

This study we derive a predictive model for the energetic proton intensities using the Cluster mission observations in the near-Earth space environment. 
We exclude the region of the radiation belts since the proton intensity levels in this region are much higher than those in the outer magnetosphere. 
Our experience has shown that the model tends to predict only the intensities in the radiation belts if it is included. 
To enable modeling the complex non-linear multidimensional dependencies, we employ a machine learning model instead of simple linear models.

To summarize, our study aims are to (1) test the capability of machine learning algorithms to predict energetic particle populations in the near-Earth space; (2) reveal which parameters are the most important for the prediction of energetic protons at different energies and (3) help future missions planning to deal best with the effects of SP.

\section{Observations and data analysis} \label{sec:data}

In this section, we describe which data we use, how it was obtained and preprocessed.

\begin{table}
    \centering
    \begin{tabular}{lll}
    \textbf{Feature} & \textbf{Unit} & \textbf{Description} \\
    \hline
    \texttt{x, y, z} & $R_E$ & position of Cluster in Geocentric Solar Ecliptic (GSE) coordinate system\\
    \texttt{rdist} & $R_E$ & radial distance from the Earth\\
    \texttt{BimfxGSE, BimfyGSE, BimfzGSE} & nT & x, y and z components of the interplanetary magnetic field (IMF) in GSE\\
    \texttt{VxSW\_GSE, VySW\_GSE, VzSW\_GSE} & km/s & x, y and z components of the SW speed in GSE\\
    \texttt{NpSW} & n/cc & solar wind density\\
    \texttt{Temp} & K & solar wind temperature\\
    \texttt{Pdyn} & nPa & solar wind dynamic pressure\\
    \texttt{AE\_Index} & nT & auroral electrojet index\\
    \texttt{SYM-H\_index} & nT & symmetric H-component index\\
    \texttt{F107} & sfu & the solar radio flux at 10.7 cm\\
    \hline
    \end{tabular}
    \caption{Overview of used input features with their units.}
    \label{tab:overview_features}
\end{table}

\subsection{Proton observations}
The Research with Adaptive Particle Imaging Detectors (RAPID) instrument~\citep{Wilken01} on four Cluster satellites~\citep{Escoubet01} measures distributions of energetic electron and ion intensities from $\sim$30 keV to $\sim$4 MeV. 
Around 50 data products are produced from the raw data and delivered to the Cluster Science Archive (CSA)\footnote{https://csa.esac.esa.int} by the RAPID team. 
We chose to work with proton observations from spacecraft (SC) 4 (Tango), which has continuous observations from 2001 through the present day. 
We can combine data from this spacecraft with proton observations by the Cluster Ion Spectrometry (CIS) instrument~\citep{Reme01}. 
The omnidirectional energetic proton intensities can be found at CSA under the product \texttt{proton\_Dif\_flux\_\_C4\_CP\_RAP\_HSPCT}~\citep{Daly10}.
We took the first seven energy channels as the labels in our experiment, which represented the energy ranges p1=28--64 keV, p2=75--92 keV, p3=92--160 keV, p4=160--374 keV, p5=374--962 keV, p6=962--1885 keV, p7=1885--4007 keV, respectively.
% TODO: We only use channels 1 to 5
We exclude the region of the radiation belts (radial distance, \texttt{rdist}$<$6 R$_\mathrm{E}$) from the dataset.

For our experiments, we preprocessed the data as follows. 
First, we eliminated outliers, namely values below 0 (NAN values) and above 10$^8$. Then, we sampled the data from their original 4-second resolution down to 1 minute because we have observed many rapid fluctuations within each minute, and the predictors related to the solar and geomagnetic activity have the highest resolution of 1 minute. 
To be more precise, we calculated the mean proton intensities each minute for each energy channel and used the beginning of each minute as a timestamp.

Since the proton intensity values span multiple orders of magnitude,  we use the common logarithms of the intensities as input to the model. However, since the proton intensity measurements contain zero values, we cannot directly apply the logarithm.
We investigated two methods:
replacing the zero values with very small values, i.e., one-tenth of the smallest non-zero value, or dropping all measurements where the intensity is zero.

We obtained better results when dropping the zero values in the sense that, in this case, the model was more focused on predicting values above zero. In considering zero values, the model was skewed to the prediction of zero values as they are many. Since high proton intensities are dangerous for the performance of the X-Ray telescopes, we have decided to base our model on the proton intensities above zero values. Developing of the model that focuses on the zero/or close to zero proton intensities requires a separate study. 
Since the number of zero values differs across channels, we performed these operations independently for each channel. 
The highest energy channels p6 and p7 were dropped from the dataset because they contain too many missing and zero values.

From 15:21:00 on 9 January 2001 to 09:57:00 on 19 February 2018 UT, 6,051,937 minutes of data in total matched these criteria. We list the predictors in Table 1. Their distributions, along with the distribution of the proton intensities, are shown in Figure~\ref{fig:hist} in the Appendix.

\subsection{Predictors}
In this subsection we introduce the predictors that we divide into groups: related to location in space and related to the solar, solar wind, and geomagnetic activity.
\subsubsection{Location}
\begin{figure}
\centering
\includegraphics[width=\textwidth]{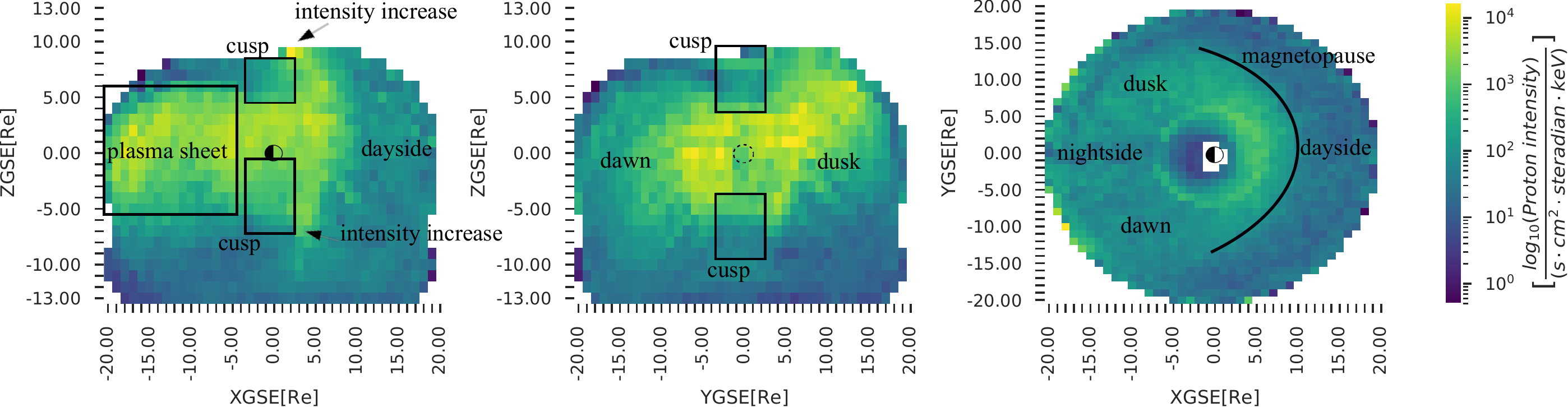}
\caption{
Distribution of the observed proton intensity for the energy channel p1 by SC4 from January 2001 to February 2018 in the GSE coordinate system. Resolution (bin size) is 1 R$_\mathrm{E}$. Half black and half white circles, as well as dashed circle, indicate the location of the Earth and are not to scale. Here zero values of proton intensity are replaced with small values.
\label{fig:distrnum}
}
\end{figure}

\begin{figure}
    \centering
    \includegraphics[width=\textwidth]{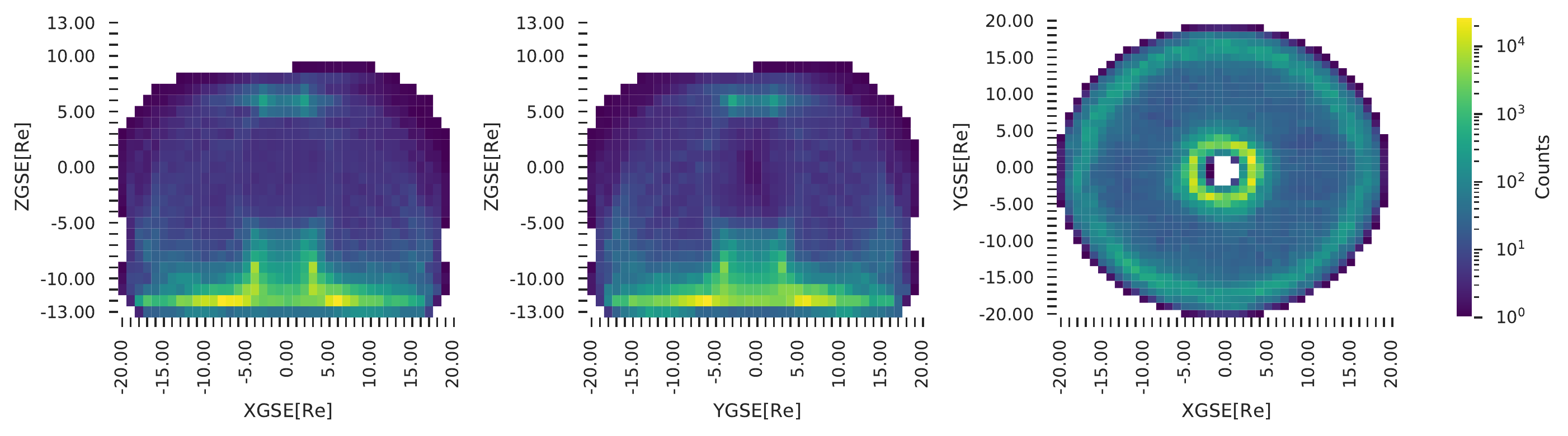}
    \caption{
    Distribution of the number of proton intensity measurements for the energy channel p1 in the GSE coordinate system in the same format as Figure \ref{fig:distrnum}.  Resolution (bin size) is 1 R$_\mathrm{E}$. 
    }
   \label{fig:distr}
\end{figure}

Each 1 minute averaged value of proton intensity is associated with a location in the Geocentric Solar Ecliptic (GSE) coordinate system represented by parameters \texttt{x}, \texttt{y} and \texttt{z}. 
The position coordinates were taken from the \texttt{sc\_r\_xyz\_gse\_\_C4\_CP\_AUX\_POSGSE\_1M} auxiliary dataset for SC4. 
We use them to obtain the radial distance from the Earth: the parameter \texttt{rdist}. 
Throughout the paper, distances are given in R$_\mathrm{E}$ units. 
The distribution of the proton intensities and the number of their samples in the GSE system is shown in Figures~\ref{fig:distrnum} and \ref{fig:distr}. 

Figure~\ref{fig:distrnum} (left and middle panels) shows high intensities around the equatorial plane, ZGSE=0, namely the plasma sheet at the night side and the region on closed magnetic field lines at the dayside in the XZ and YZ planes. 
There are also mid-altitude cusps with less intensive plasma around XGSE=0, in the same planes, above and below $\sim$3R$_\mathrm{E}$ and $\sim$-3R$_\mathrm{E}$ in ZGSE direction, respectively. 
Enhanced proton intensities at XGSE$\simeq$3-6R$_\mathrm{E}$ at higher latitudes than the plasma sheet (above and below $\sim$5R$_\mathrm{E}$ and $\sim$-5R$_\mathrm{E}$ in ZGSE direction, respectively) in the XZ plane are also visible. 
The intensities are higher in the northern hemisphere than in the southern hemisphere. 
The dusk-dawn distributions (in the YZ plane) show asymmetry with higher intensities at the northern hemisphere's dusk side.
The same higher intensity at dusk is visible in the XY plane.
This asymmetry agrees with observations of energetic protons $>$274 keV by~\citet{Kron15,Luo17}. 
The proton intensities are higher at the dayside in the region on closed magnetic field lines. 

\begin{figure}
	\includegraphics[width=1\linewidth]{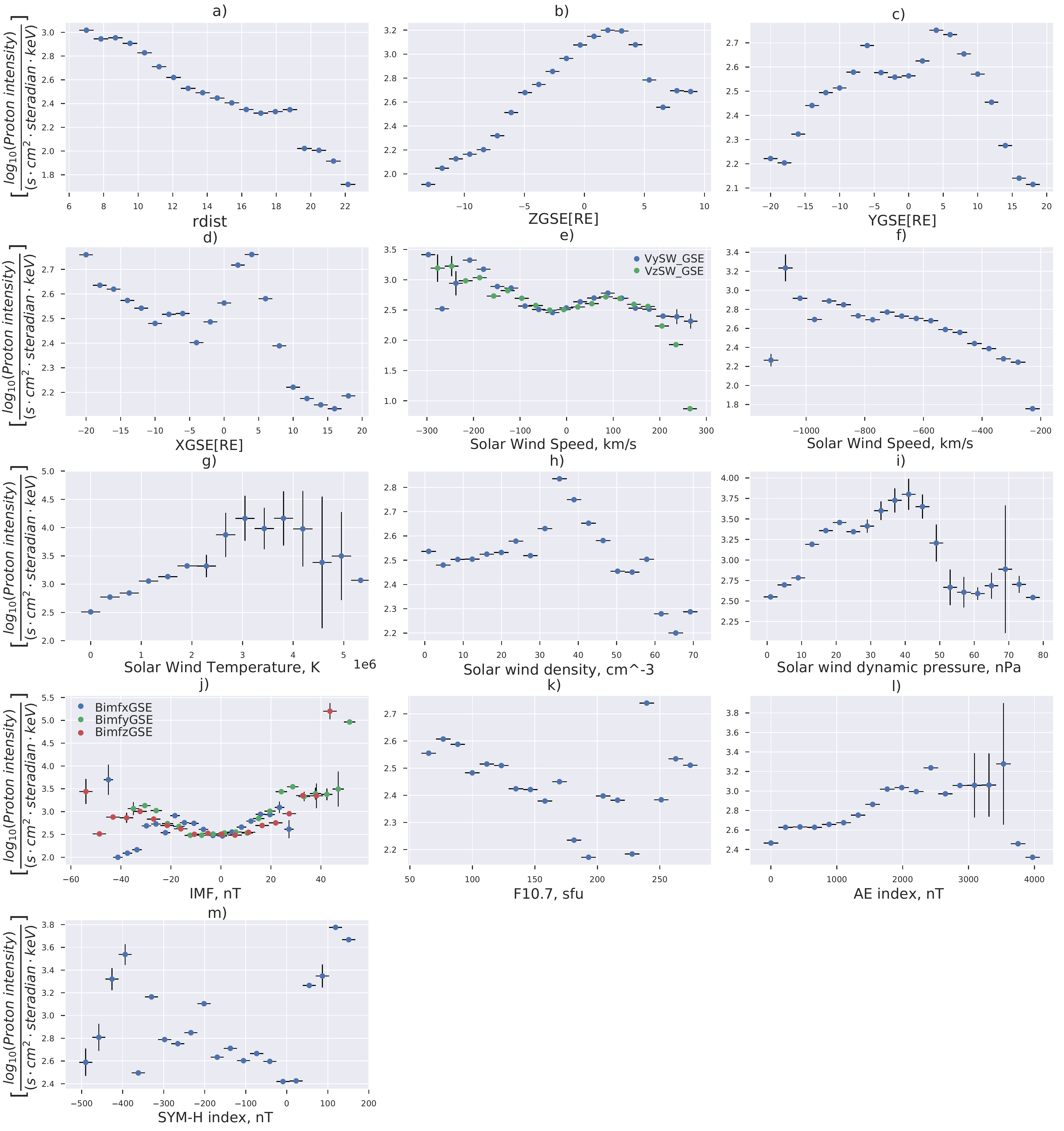}
	\caption{
	Relations of mean proton intensities for the energy channel 1 and (a)-(d) radial distance, \texttt{rdist}, ZGSE, YGSE, and XGSE, respectively; (e) the solar wind $V_y$ and $V_z$ components; (f) the solar wind radial velocity, $V_x$; (g)-(i) SW temperature, density, and dynamic pressure, respectively; (j) IMF components in GSE; (k) F10.7 parameter; (l) AE index and (m) SYM-H index. Vertical lines represent confidence intervals at 95\% confidence level. Here zero values of proton intensity are replaced with small values.
	\label{fig:binned}
	}
\end{figure}
In Figure~\ref{fig:binned}, we plot the mean proton intensities versus individual predictors. 
In the following, for brevity, we refer to the logarithm of proton intensities simply as proton intensities.
In panel (a), we observe a strong, almost linear decrease of the proton intensities with the radial distance.
The intensities almost linearly decrease with the ZGSE  coordinate in the southern hemisphere (see panel (b)), while the dependence is more complicated in the northern hemisphere.
Panel (c) shows that the maximum of the proton distribution is around $\pm$5 R$_\mathrm{E}$. 
The proton intensity falls rapidly at $\sim$-15 R$_\mathrm{E}$ at the dawn side and $\sim$13 R$_\mathrm{E}$ at the dusk side. 
The dependence of proton intensities on XGSE is rather complicated, see panel (d). The strong increase of the proton intensities at XGSE$<$-15 R$_\mathrm{E}$ can be explained by reduced spacecraft sampling in the northern lobe at this distances, see XZGSE plane in Figure \ref{fig:distr}.
It falls strongly at $\sim$10 R$_\mathrm{E}$ as expected at the magnetopause boundary, as also seen in Figure~\ref{fig:distrnum}.
These spatial dependencies resemble those observed for the SP contamination at the XMM-Newton X-Ray telescope in~\citet{Kron20a}, see also~\citet{Ghizzardi:2017}.

\subsubsection{Solar, solar wind and geomagnetic activity}

We combine the proton intensities with simultaneous observations of solar, SW, and geomagnetic parameters taken from the OMNI database\footnote{\url{https://omniweb.sci.gsfc.nasa.gov}}; see also~\citet{Papitashvili05}. 
The SW observations are propagated to the Earth's bow shock. 
The SW is characterized by the proton density, NpSW in cm$^{-3}$; components of the speed (VSW) in the GSE coordinates, \texttt{VxSW\_GSE}, \texttt{VySW\_GSE} and \texttt{VzSW\_GSE} in km s$^{-1}$; the temperature, \texttt{Temp}, in K; the dynamic pressure, \texttt{Pdyn} in nPa, which is calculated as NpSW*VSW$^2\times1.67\cdot 10^6$, and components of the IMF in the GSE coordinates, \texttt{BimfxGSE},  \texttt{BimfyGSE} and  \texttt{BimfzGSE}, in nT. 

The SW velocity $V_y$ and $V_z$ components show that the deviation from the radial direction leads to an increase of the proton intensities, excluding cases with substantial deviation ($>$100 km/s) in positive Y- and ZGSE directions for which a decrease is observed, see Figure~\ref{fig:binned} (e). 
The proton intensities increase with the SW speed in the anti-sunward direction, $V_x$, and the temperature, see panels (f) and (g). 
The dependencies of the proton intensities on SW density (panel (h)) and the SW dynamic pressure (panel (i)) have non-linear trends. The change of the IMF's direction towards stronger absolute values mainly leads to increased proton intensities (see panel (j). 

The influence of solar irradiation is investigated using the F10.7 index, which measures the radio flux at 10.7 cm (2.8 GHz;~\citet{Tapping13}). This parameter correlates well with the number of sunspots and other indicators of solar and ultraviolet solar irradiance. 
It can be measured reliably under any terrestrial weather conditions (unlike many other solar indices). It is denoted here by \texttt{F107}, and it is measured in solar flux units (sfu). The solar irradiance is non-linearly related to the proton intensities; see Figure~\ref{fig:binned} (k). 

A parameter of geomagnetic activity such as the auroral electrojet (AE) index, denoted as \texttt{AE\_index}, in nT, characterizes the magnetic field disturbance in the auroral region of the northern hemisphere~\citep{AEindex}. 
The relation of the proton intensities with the AE index is also non-linear; see Figure~\ref{fig:binned} (l). 
However, a general trend of increase of the proton intensities with the AE index is visible, namely with the geomagnetic activity at high latitudes.
Another parameter related to the geomagnetic activity is the SYM-H index, denoted as \texttt{SYM-H} and measured in nT~\citep{AEindex}. 
This parameter characterizes the disturbance of the geomagnetic field at the equatorial regions. 
The geomagnetic activity related to the geomagnetic storms, characterized by the SYM-H index, shows non-linear relation with proton intensities, see Figure \ref{fig:binned} (m).  

If we compare the trends of the proton intensity changes with the solar, SW, and geomagnetic parameters with those for the SP contamination from~\citet{Kron20a} we notice general agreement between those. 
More details on the comparison are in Section~\ref{sec:discussion}.

\subsection{Cross-correlations between proton intensities and predictors}
\begin{figure}
    \plotone{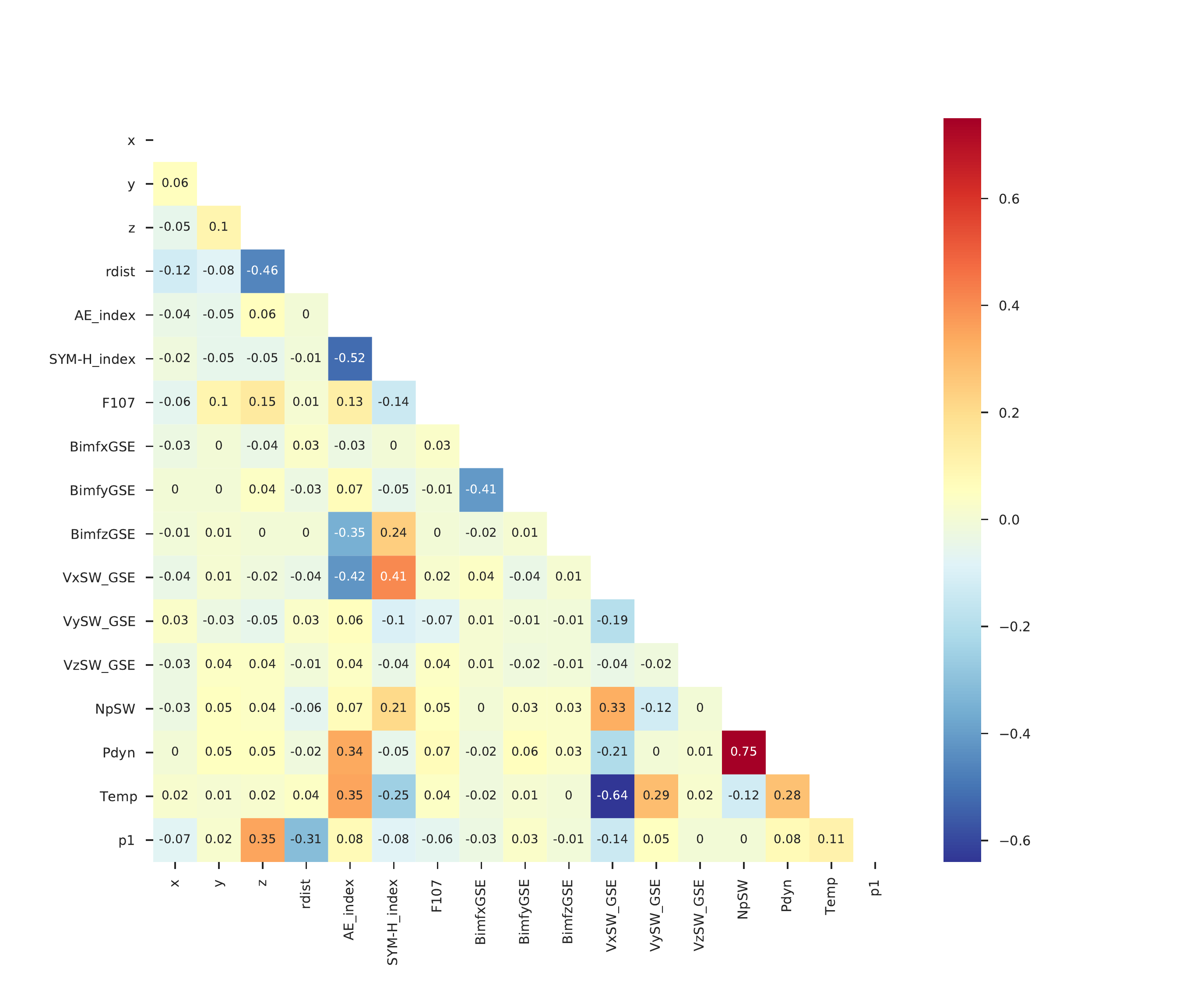}
    \caption{
    PC matrix between input parameters and proton intensity (here, zero values are replaced with small values). For readability, the values are rounded to the second decimal. 
    }
    \label{fig:corr}
\end{figure}

Figure~\ref{fig:corr} shows the Pearson Correlation (PC) between parameters possibly related to proton dynamics. The correlation values vary between -1 and 1, with values close to -1/1 meaning perfect linear anticorrelation/correlation and values close to 0 meaning no linear correlation. One should be careful with the interpretation of the PC coefficient, as this indicates only linear relationships. The proton intensities from channel p1 are well correlated with the radial distance and the direction Z, in agreement with Figure~\ref{fig:binned}.  From the OMNI parameters, the proton intensities of channel 1 are best linearly correlated with the \texttt{VxSW\_GSE}, the same as the SP contamination in~\citet{Kron20a}.

\subsection{Data Split}
The full dataset, as was mentioned above,  comprises in total 6,051,937 measurements from 2001-01-09 15:21:00 UT to 2018-02-19 09:57:00 UT.
We split the dataset into a training (or development) set (80\%) and a test set (20\%).
To prevent test leakage, we do not shuffle the data but split it by a time point with the original order preserved.
Afterward, we additionally split the development data into a train (80\%) and validation set (20\%) again by time.
We utilize the validation set to optimize the model hyperparameters.  

We normalized the features by subtracting the median, and dividing by the inter-quartile range.\footnote{As implemented by \texttt{RobustScaler} in \texttt{sklearn}}

\begin{table}
\caption{
Size and periods of the data subsets after splitting. 
The time is given in UT.
}
    \label{tab:split}
    \centering
    \begin{tabular}{*{3}{l}*{5}{r}}
    \toprule
    Subset & Start & End & \multicolumn{5}{c}{Number of Data Points without Zero Values}\\
    &&&p1&p2&p3&p4&p5     \\
    \midrule
    Train & 2001-01-09 15:21:00 & 2011-08-23 08:22:00 & 2,130,927 & 978,828 & 1,396,154 & 1,093,985 & 522,122\\
    Validation & 2011-08-23 08:24:00 & 2014-07-24 22:44:00 & 532,731 & 244,707 & 349,038 & 273,496 & 130,530\\
    Test & 2014-07-24 22:45:00 & 2018-02-19 09:57:00 & 694,867 & 320,363 & 447,790 & 329,581 & 131,208\\
    \bottomrule
    \end{tabular}
\end{table}

Table~\ref{tab:split} summarizes the sizes and periods of the data subsets after performing preprocessing and splitting.

\section{Machine learning model for Proton Intensities}
\label{sec:model}

The relation between the proton intensities and the different predictors listed above is complex; see Figure~\ref{fig:binned}.
It is, therefore, often a group of predictors or their ensemble that leads to better predictions than the best individual predictor~\citep{geron19}. 

We interpret the prediction of proton intensities as a regression problem of form
\[
y \approx f([\mathbf{p}; \mathbf{x}_t] \mid \mathbf{\theta})
\]
where $y$ is the proton intensity for a single channel, $\mathbf{p}$ is the spatial position, and $\mathbf{x}_t$ the additional input parameters as given in Table~\ref{tab:overview_features} at time $t$.
$[~;~]$ denotes the concatenation, and $\theta$ are model parameters that are to be estimated from the given dataset.
We study solving this regression task by applying various machine learning models, from which we select according to validation performance.
The models' hyperparameters are also tuned according to validation performance.

We investigated the following linear models:
linear regression~\citep{Galton1886} with and without $l_1$/$l_2$ regularization (lasso~\citep{Santosa1986}/ridge regression~\citep{Hoerl1970}), least angle regression~\citep{Tibshirani2004} with and without $l_1$ regularization (lasso LARS regression), and linear support vector regression (SVR)~\citep{Cortes1995}.

We considered the following Tree-based Ensemble models:
regression trees~\citep{breiman1984}, random forests, AdaBoost~\citep{Freund1995}, ensembles of extremely randomized trees (extra trees)~\citep{Geurts2006}, gradient boosting~\citep{Friedman2001}, and light gradient boosting machines (LGBM)~\citep{Ke2017}.

We also used a multi-layer perceptron (MLP) regressor~\citep{rosenblatt1958perceptron}. 
This regressor is a neural network with an input layer, one or more hidden layers, and one output layer.
Each layer is a dense layer, $\mathbf{x}^{(i+1)} = \sigma(\mathbf{W}\mathbf{x}^{(i)} + \mathbf{b})$, where $\mathbf{W} \in \mathbb{R}^{d_i \times d_{i+1}}, \mathbf{b} \in \mathbb{R}^{d_{i+1}}$ are trainable parameters, and $\sigma$ is a non-linear activation function.
The weights are trained \emph{end-to-end} with an iterative (stochastic) gradient descent method using the gradient of the loss function's output w.r.t. parameters.
The gradients can be efficiently computed using back-propagation.
Important hyperparameters are the choice of the network structure, i.e., the number of layers and the hidden dimensions $d_i$.

\section{Machine Learning Experiments}

\subsection{Setup}
We used LGBM implementation from LightGBM library~\citep{lgbm} and  sklearn~\citep{sklearn_api} for all other models.
In order to evaluate the models' performance, we use the following measures:
Spearman Correlation (SC), PC, Mean Absolute Error (MAE), Mean Squared Error (MSE), and coefficient of determination (R$^2$).
For model selection and hyperparameter optimization, we focus on SC.

To demonstrate the necessity of using advanced machine learning models, we compare the performance against two simple baselines: historical binning and k-Nearest Neighbors (kNN)~\citep{knn}.

In the historical binning baseline, we create spatial bins of training data with the k-means algorithm~\citep{lloydleast} applied to the position features only. 
The number of bins is chosen independently for each channel based on validation performance.
For a test point, we determine the corresponding bin and use the average proton intensity of that corresponding region from the train data as the prediction.

In the kNN baseline, for a test data point, we determine the k nearest neighbor that exists in the train set based on the Minkowski distance and interpolate its proton intensity for the position. 
Then we take the interpolated intensity at the test position as the prediction.

\subsection{Results}
\label{sec:modeltrain}
\begin{table}
\centering
\caption{
SC between predicted by different models and observed proton intensity for all channels on both train and test split. %, both, validation and test split.
The MLP model was chosen as the best configuration according to the validation SC of channel p1.
}
\label{tab:hpo-table}

\begin{tabular}{lrrrrrr}
\toprule
Model & Split &     p1 &     p2 &     p3 &     p4 &     p5 \\
\midrule
\midrule

kNN   & Test           & 29.05\% & 32.32\% & 32.34\% & 32.74\% & 29.98\% \\
HistBin  & Test       & 39.30\% & 44.04\% & 40.28\% & 42.33\% & 43.99\% \\
\midrule
MLP     & Train        & 73.48\% & 72.62\% & 70.58\% & 70.83\% & 74.84\% \\
MLP     & Test        & 56.89\% & 58.82\% & 53.47\% & 56.15\% & 53.49\% \\
\bottomrule
\end{tabular}
\end{table}

In order to determine the best hyperparameters, we used random search over a pre-defined search space (see Table~\ref{tab:hpo} in the Appendix) due to its higher sample efficiency than grid search~\citep{bergstra2012}.
In this search, we trained numerous models on the training data.
We utilized the ASHA Scheduler~\citep{li2020massively} for its parallelism and extensive early stopping capabilities via the \texttt{ray tune} \citep{liaw2018tune} framework.
After training, we evaluate each model on the validation data, and we choose the configuration which obtains the best validation SC.
To reduce resource consumption, we perform this optimization only for channel p1 and apply the same best hyperparameters for the other channels, cf. Table~\ref{tab:hpo} in the Appendix.\footnote{We publish the trained models at \url{http://doi.org/10.5281/zenodo.4593065}}
We obtain the best validation performance for an MLP model.
Table~\ref{tab:hpo-table} shows the final results on the test set. 
The MLP model outperforms the baseline models such as kNN and historical binning by about 80\% and 33\%, respectively.
The average SC correlation between the observed and predicted data is about 56\%. 
This value is reasonable considering the stochasticity and complex dynamics of the fast-moving energetic protons in the terrestrial magnetosphere.

\begin{figure}
  \centering
  \includegraphics[width=\linewidth, height=0.2\textheight, keepaspectratio]{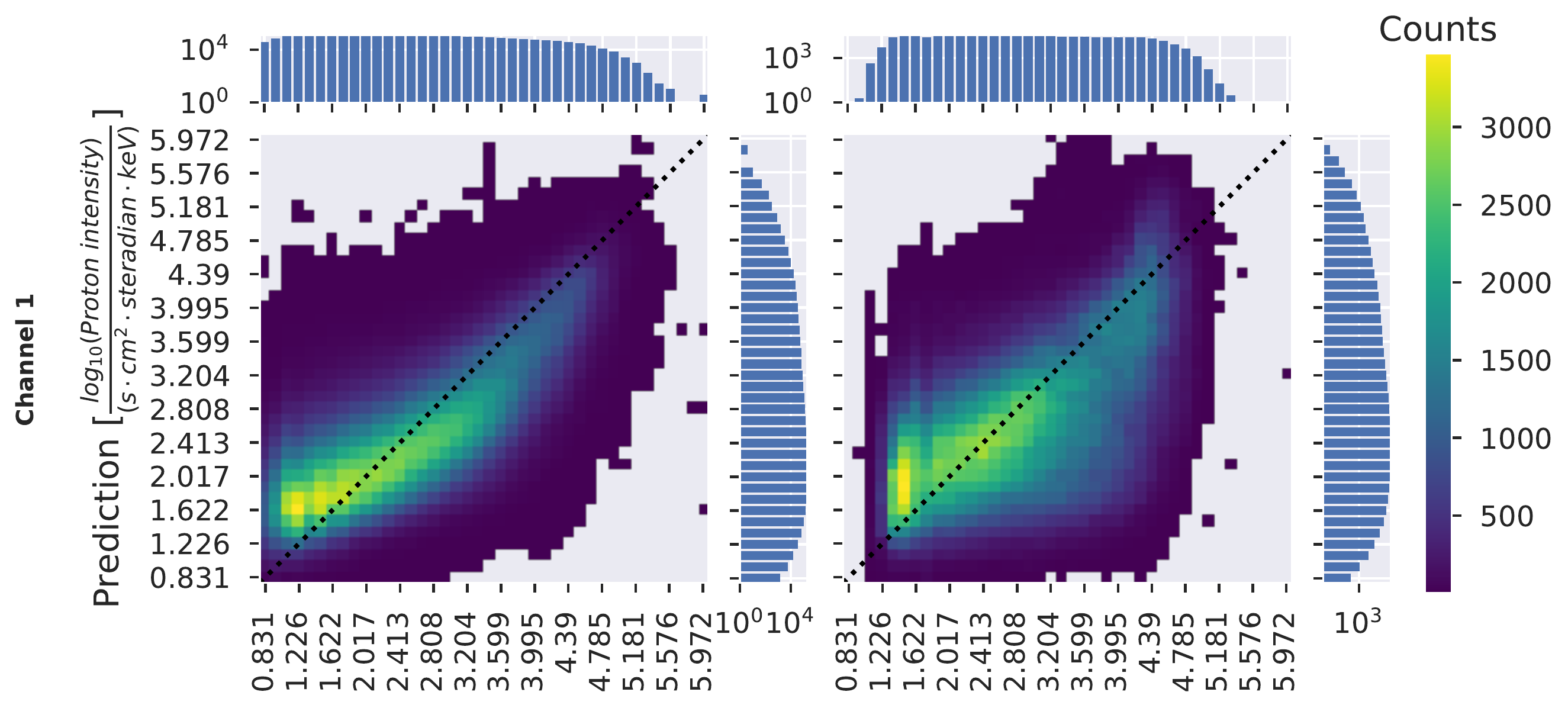}
  \includegraphics[width=\linewidth, height=0.2\textheight, keepaspectratio]{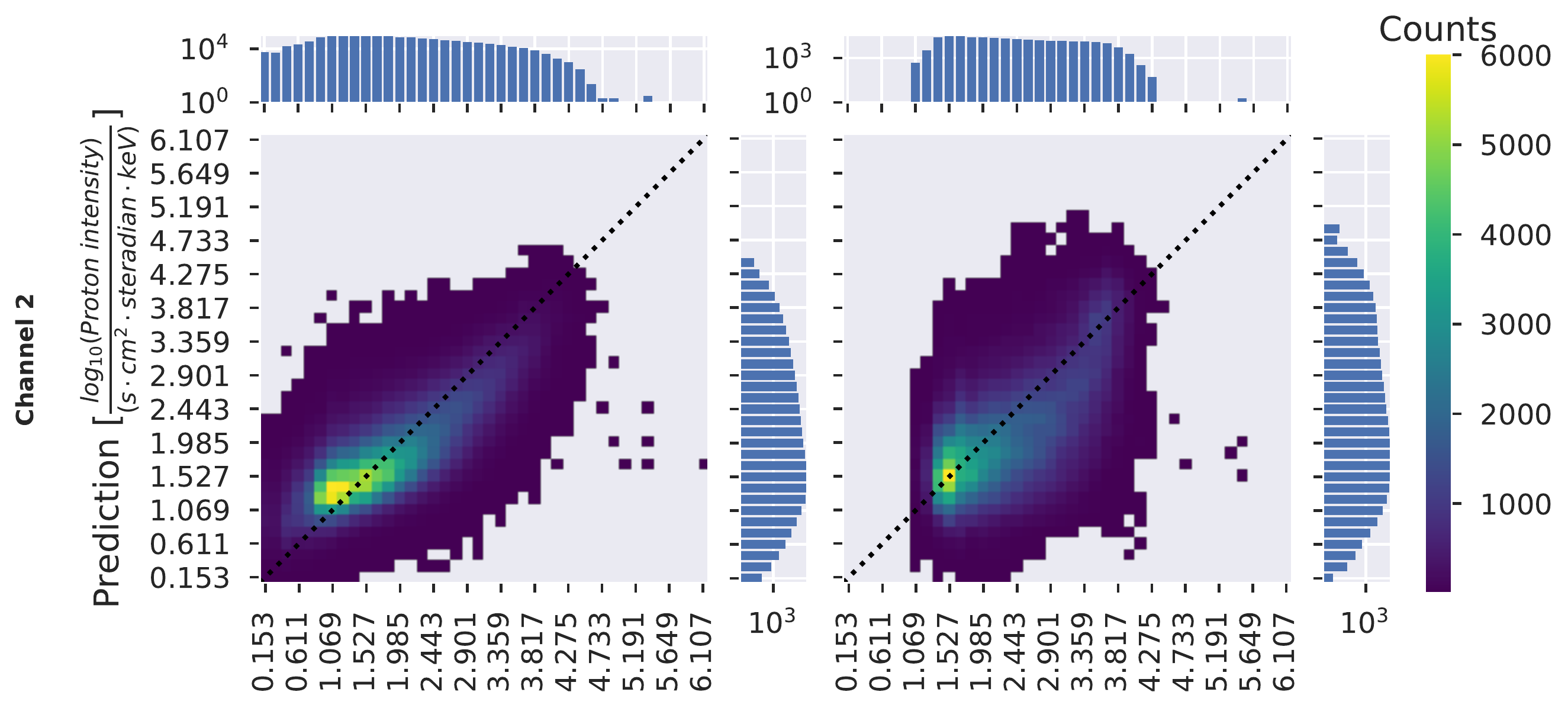}
  \includegraphics[width=\linewidth, height=0.2\textheight, keepaspectratio]{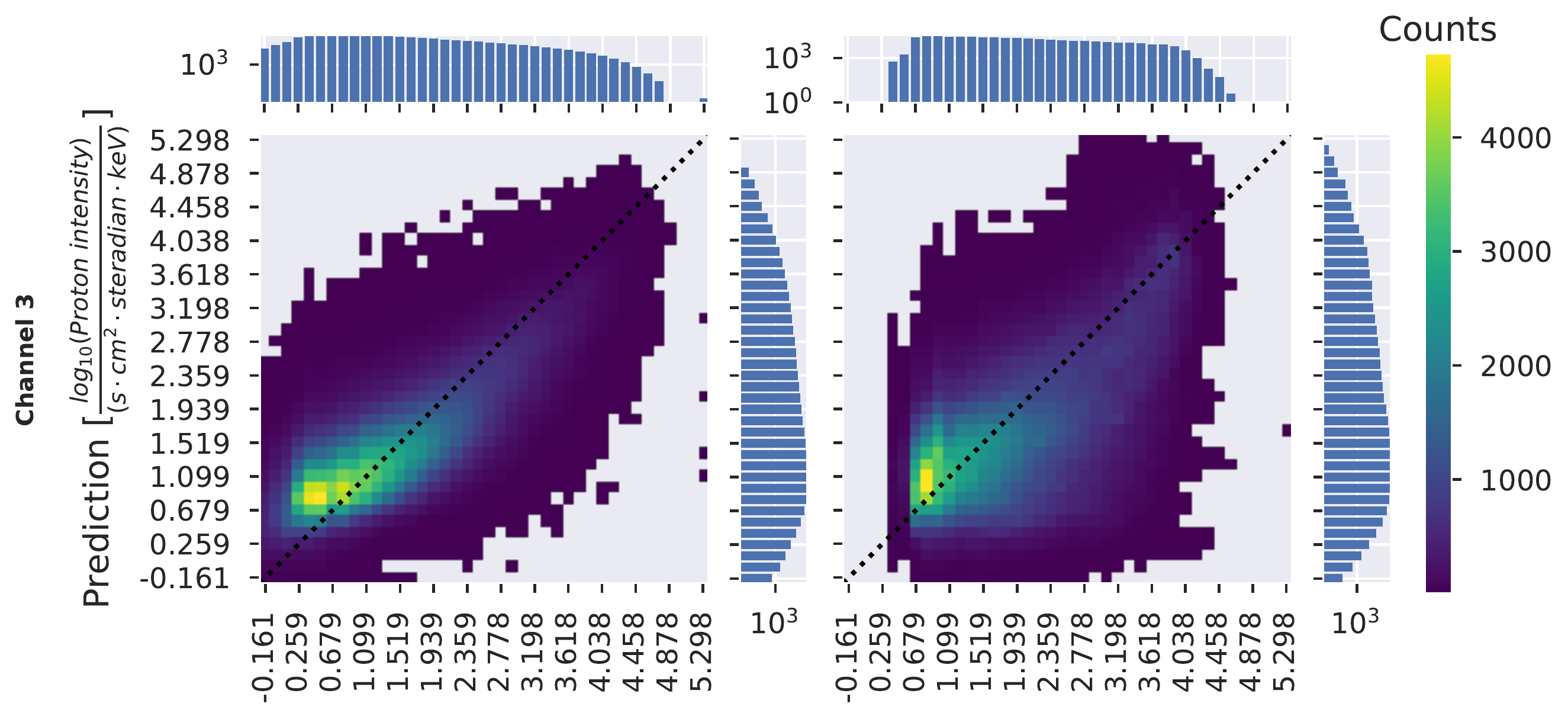}
  \includegraphics[width=\linewidth, height=0.2\textheight, keepaspectratio]{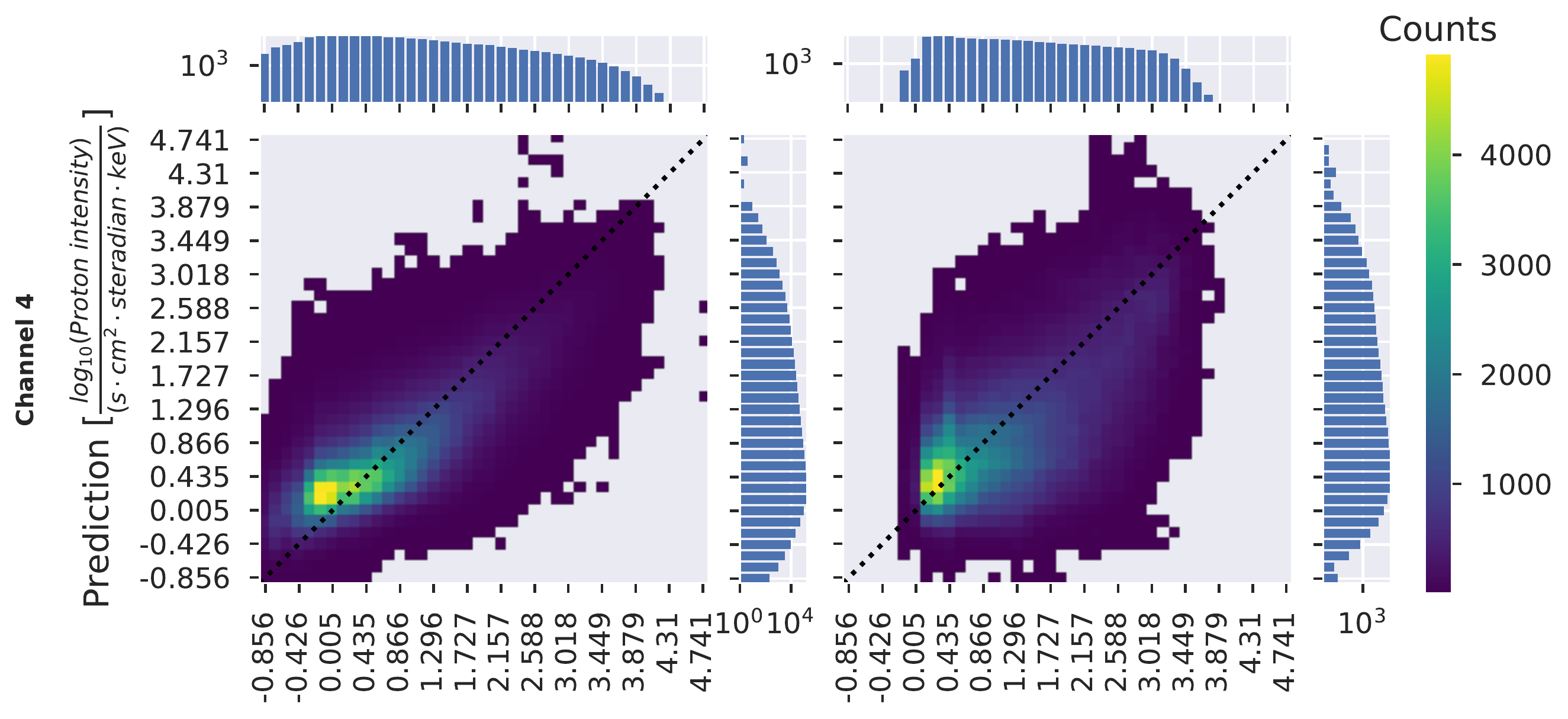}
  \includegraphics[width=\linewidth, height=0.2\textheight, keepaspectratio]{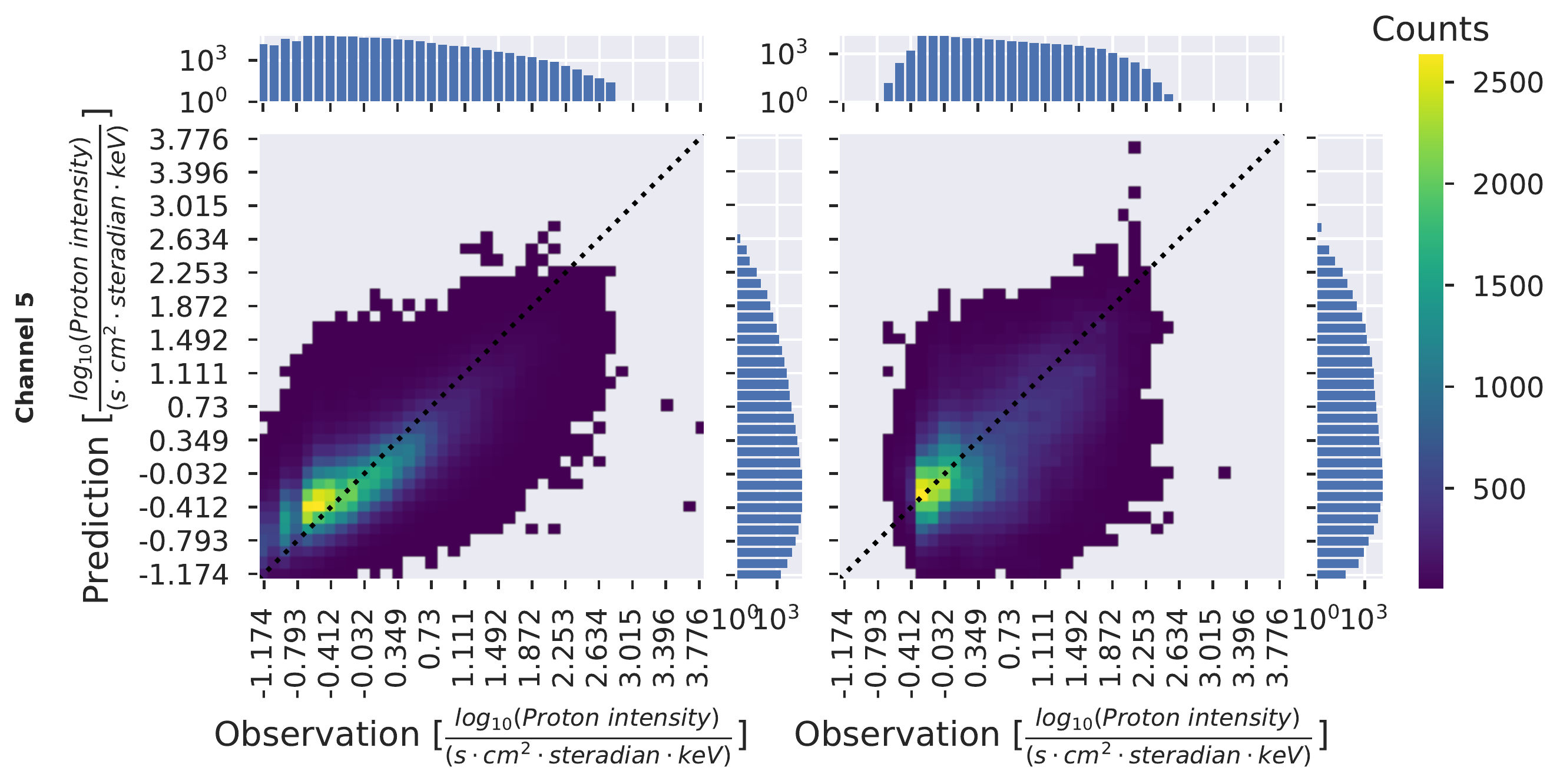}
\caption{
The observed (x-axis, in this case zero values are dropped) vs. the predicted (y-axis) proton intensities for the energy channels p1 to p5 from the training (left) and the test (right) set. 
The color represents the number of samples in the corresponding bin, where a good model will predict most of the intensities along the diagonal, i.e., closely matching the measurements. 
The underlying model for those plots is the best MLP determined according to validation SC for channel p1.
}
\label{fig:predicted-distribution}
\end{figure}
Figure~\ref{fig:predicted-distribution} shows the distributions of the observed proton intensities versus the predicted values for five energy channels.
The left-hand panels show the training set distributions, whereas the right-hand panels represent the test set. Observed and predicted data for the training and test data sets agree relatively well.
The data is mainly concentrated along the black dashed line, corresponding to a perfect correlation. 
The correlations are not much different between the training and test set, implying that the training data was not over-fitted.
The prediction of proton intensities in all five energy channels has a similar quality. 

\begin{figure}
    \centering
	\includegraphics[width=0.8\linewidth]{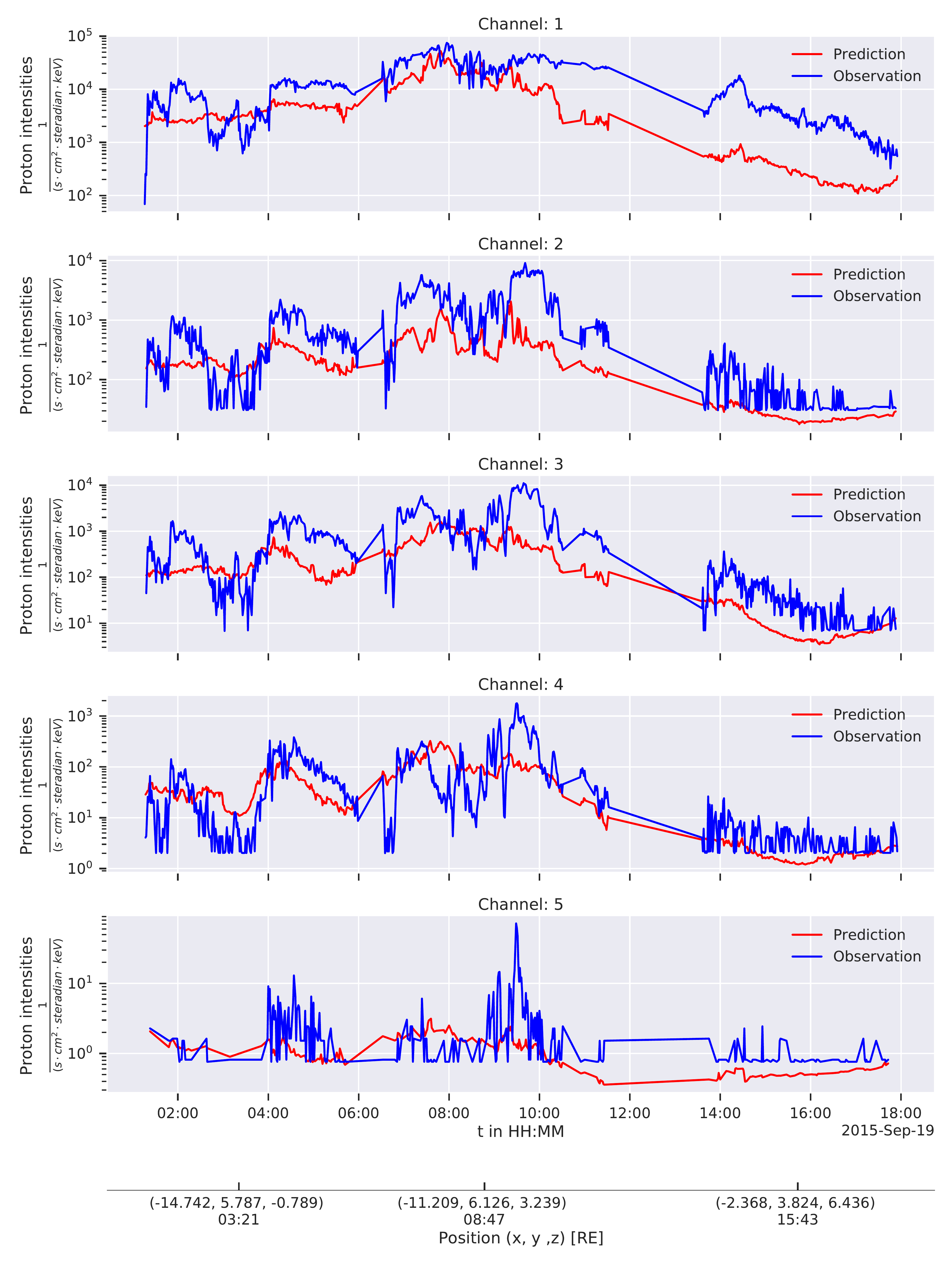}
	\caption{Predicted intensities (red) vs. measured intensities (blue, in this case, zero values are dropped) within the time interval on 2015-09-19 from 01:15 to 20:30 UT.  \label{fig:example-prediction-test}}
\end{figure}

In Figure~\ref{fig:example-prediction-test}, we show a qualitative example of the model's predictions. 
The chosen time frame shows a plasma sheet crossing, the approach of the radiation belts, and the cusp region at the nightside of the magnetosphere. The model predicts the proton intensities for the different energy channels, rarely deviating more than one order of magnitude, which is considered a good prediction for energetic particle dynamics.

To evaluate the statistical significance of our results, we use a 2-sided hypothesis test on the SC.
The null hypothesis states that the predictions are uncorrelated to the observations.
For all five channels, we obtain a $p$ value of $0$, or, more correctly, below the smallest positive number in \texttt{float64}.
Hence, we can reject the null hypothesis, i.e., the predictions are correlated to the observations.
Thus, we can deduce from this result that our model learned the trend of proton intensity.

\subsection{Feature Importances}
\begin{figure}
    \centering
	\includegraphics[width=\textwidth]{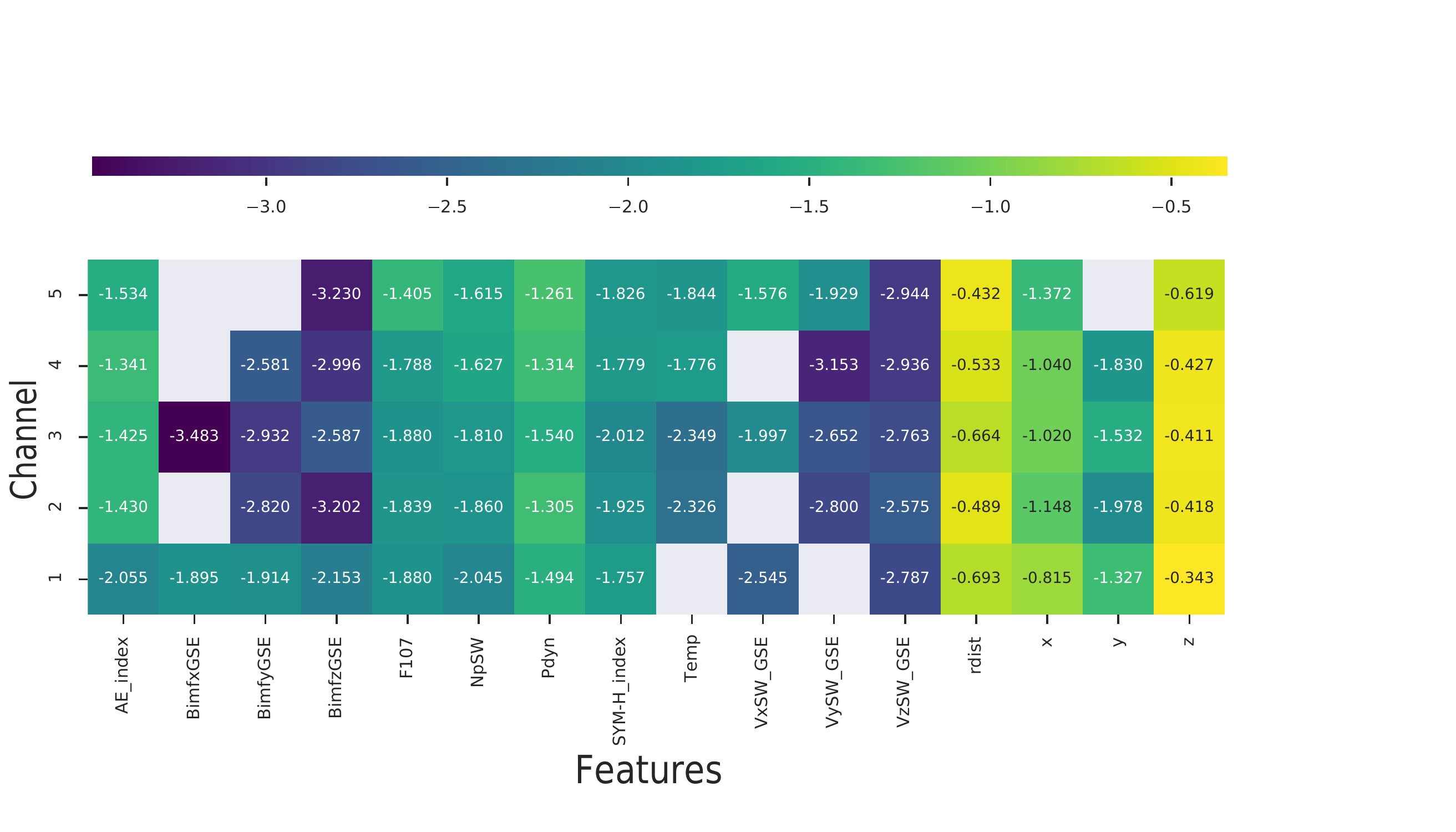}
	\caption{Permutation feature importance in the logarithmic (base 10) scale calculated for the MLP model on the test set. Original negative values of the feature importance are shown in gray color.}
	\label{fig:importance}
\end{figure}
To investigate the importance of the individual input features on the model's prediction, we utilize permutation feature importance~\citep{fisher2019models} since it is model-agnostic and interpretable.
For an investigated feature column, e.g., $x$, its values are shuffled to break any association with the target value, i.e., the proton intensity. 
Then, we re-calculate the model's prediction and evaluate the SC between the prediction and the original target values.
The feature importance is measured as the decrease in performance: thus,  a more substantial decrease in model performance corresponds to a more important feature.
Negative feature importance is also possible when the prediction improves when breaking an input feature's association.
We use bootstrapping with 100 iterations and a subsampling fraction of 75\% to estimate the mean and variance of this statistic to assess the significance.

Figure~\ref{fig:importance} shows the feature importance for each input variable and channel. The parameters related to the location show significantly higher importance than parameters related to solar, SW, and geomagnetic activities. 
From those, on average, the most vital dependence is seen for ZGSE and the radial distance. 
The least important location parameter is YGSE. 
From the other parameters, the SW dynamic pressure is the most important parameter for predicting the proton intensities, followed by, on average, the AE index.

\section{Discussion}  \label{sec:discussion}
We want to note that the most substantial linear dependence of the proton intensities among the OMNI parameters is on the \texttt{VxSW\_GSE}, cf. the PC in Figure~\ref{fig:corr}. 
This linear dependence is also evident in Figure~\ref{fig:binned}. 
However, the feature importance derived in our model indicates \texttt{Pdyn} and AE index as the most important OMNI parameters, cf. Figure~\ref{fig:importance}.
In Figures~\ref{fig:binned} and \ref{fig:corr}, we always consider one input variable and one output variable in isolation. 
The model, however, is not restricted to do so, but rather can combine features to get ``latent features'', i.e., combinations of individual features or their interaction.

The proton intensities' dependency to different parameters are generally very similar to the dependencies of the XMM SP contamination count rates derived in~\citet{Kron20a}; despite differences in the trajectories of the spacecraft.
For example, a clear dependence on \texttt{rdist} is similar in enhancing the proton intensities and count rates between 17 and 19 R$_\mathrm{E}$. 
For the YGSE dependency, we observe for both proton intensities and SP count rates a decrease at YGSE$\simeq$0 R$_\mathrm{E}$ and dusk-ward asymmetry at distances YGSE$\simeq\pm$10 R$_\mathrm{E}$. 
For the XGSE dependency, we observe high proton intensities and SP count rates between XGSE$\sim$0 and 5 R$_\mathrm{E}$. 
Similar dependencies are observed for the almost linear relation with \texttt{VxSW\_GSE} up to $\sim$-900 km/s, for the \texttt{Temp}, \texttt{Np} up to 20 cm$^{-3}$, \texttt{BimfxGSE} and \texttt{BimfyGSE} up to $|$10$|$ nT, AE index up to $\sim$700 nT and the SYM-H index within the range between -150 to 50 nT.
For \texttt{Pdyn}, the considered scale ranges are very different, and, therefore, the resolution is different. However, we still see an average increase of \texttt{Pdyn} with the proton intensities/SP count rates at least up to 6 nPa. 
The only considerable disagreement for the relations of the proton intensities and parameters is observed for the F10.7 index. For example, for the XMM count rates, one observes an increase with the F10.7 index up to 150 sfu. However, the proton intensities observed by Cluster show a decrease. 
We can draw a relation to the complexity of the processes associated with the solar cycle or/and to the bias related to the spacecraft trajectory during the solar cycle. 
The Cluster mission was located closer to the Earth during the years related to the 24th solar cycle (higher proton intensities versus lower F10.7 index) that is significantly weaker than the 23rd solar cycle during which the XMM measurements were done (lower proton intensities versus higher F10.7 index). 
The highest PC derived for the proton intensities/SP count rates are the ZGSE, \texttt{rdist}, and \texttt{VxSW\_GSE} in both cases. 
In conclusion, based on Figures~\ref{fig:binned} and \ref{fig:corr} our energetic proton observations indicate that it is likely that they can be a cause of the XMM SP contamination. 
The physical processes leading to proton energization associated with high SW speed are related, for example, to the formation of the quasi-parallel shock, acceleration processes associated with reconnection at the dayside, magnetospheric storms, and substorms, see more detailed discussion in~\citet{Kron20a}.

The dependencies of proton intensities and the SP count rates on the different parameters are generally similar. 
A notable difference is that  \texttt{Pdyn} and AE index had the highest importance of the proton intensities in the MLP model. 
In contrast, the SP count rates in the Extra Random Forest model had the highest importance for the \texttt{VxSW\_GSE} and F10.7 index (from the solar, the SW and geomagnetic parameters).
This difference most likely occurs because of using different non-linear models and different observations' lengths (one vs. one and a half solar cycles). 
Another possible reason is that the Cluster trajectories covering the plasma sheet much more thorough than those of the XMM. 
The plasma sheet dynamics is strongly related to substorm activity indicated by the AE index and is strongly dependent on \texttt{Pdyn}. 
In the study by~\citet{Smirnov19} based on Cluster observations, the dependence of the electron intensities at L-shells between 4 and 6 for $\sim$1.5 solar cycle (approximately the same period as in our study) was also the best correlated with the \texttt{Pdyn} and AE index. 
The electron intensities in the outer radiation belts are governed by the substorm associated activity in the plasma sheet and acceleration processes related to charged particle injections/dipolarizations~\citep[e.g.,][]{gabrielse14,Malykhin18}. 
The proton intensities are also enhanced during such injections.  

Based on this experience, in our future study, we plan to derive a tailored machine learning model for predicting the proton intensities at different energy channels along the XMM trajectory. 
In this study, we want to use the same time interval and location in the magnetosphere to avoid possible biases. 
The aim will be to find which energies of proton intensities correlate most strongly with the SP count/rates. 
The results of the tailored model will be compared with the results of this model.

In this work, we have tried to include a wide range of the history of the solar and geomagnetic parameters. However, they did not improve the model. The SW-- magnetosphere energy coupling functions ~\citep[e.g.,][]{Gonzalez94,Milan12,Wang14} are not considered in this work and beyond the scope of this paper. 

\section{Conclusions}  \label{sec:conclusions}
Using 17 years of the Cluster/RAPID observations, we have derived a machine learning-based model for predicting the proton intensities at the energies from 28 to 1,885 keV in the 3D terrestrial magnetosphere between 6 and 22 R$_\mathrm{E}$.
As predictors, we used the location, solar, SW, and geomagnetic activity indices.
 The results demonstrate that the neural network's (MLP) prediction capabilities exceed the baselines based on the kNN and historical binning on average by $\sim$80\% and $\sim$33\%, respectively. 
The average correlation between the observed and predicted data is about 56\% despite the complex dynamics of the energetic protons in the magnetosphere. 
The most important parameters for predicting proton intensities in our model are the ZGSE direction and the radial distance, both related to location. 
The most important predictor of solar, SW, and geomagnetic activity is the SW dynamic pressure. The results are in general agreement with the study by~\citet{Kron20a} on the characteristics of the SP contamination in the XMM-Newton telescope.  The results can directly be applied in practice, e.g., to assess the contamination of the X-Ray telescopes as well as better determining the contamination risk for various future mission concept orbits.

\acknowledgments

The authors are thankful to the Cluster Science Archive (\url{https://csa.esac.esa.int}) team for providing the data. We acknowledge the use of NASA/GSFC's Space Physics Data Facility's OMNIWeb service and OMNI data. This work was conceived within the team led by Fabio Gastaldello on ``Soft Protons in the Magnetosphere focused by X-ray Telescopes'' at the International Space Science Institute in Bern, Switzerland.  EK is supported by the German Research Foundation (DFG) under number KR 4375/2-1 within SPP ``Dynamic Earth''. NS is supported by NASA Earth and Space Science Grant 80NSSC17K0433.
The computational infrastructure was provided by the Leibniz Supercomputing Centre (LRZ).

\software{sklearn \citep{scikit-learn},
       scipy  \citep{2020SciPy-NMeth},
          numpy  \citep{vanderWalt11},
          pandas  \citep{mckinney-proc-scipy-2010},
          Matplotlib  \citep{Hunter07}
          }

%\bibliography{main}{}

\begin{figure}[ht!]
  	\plotone{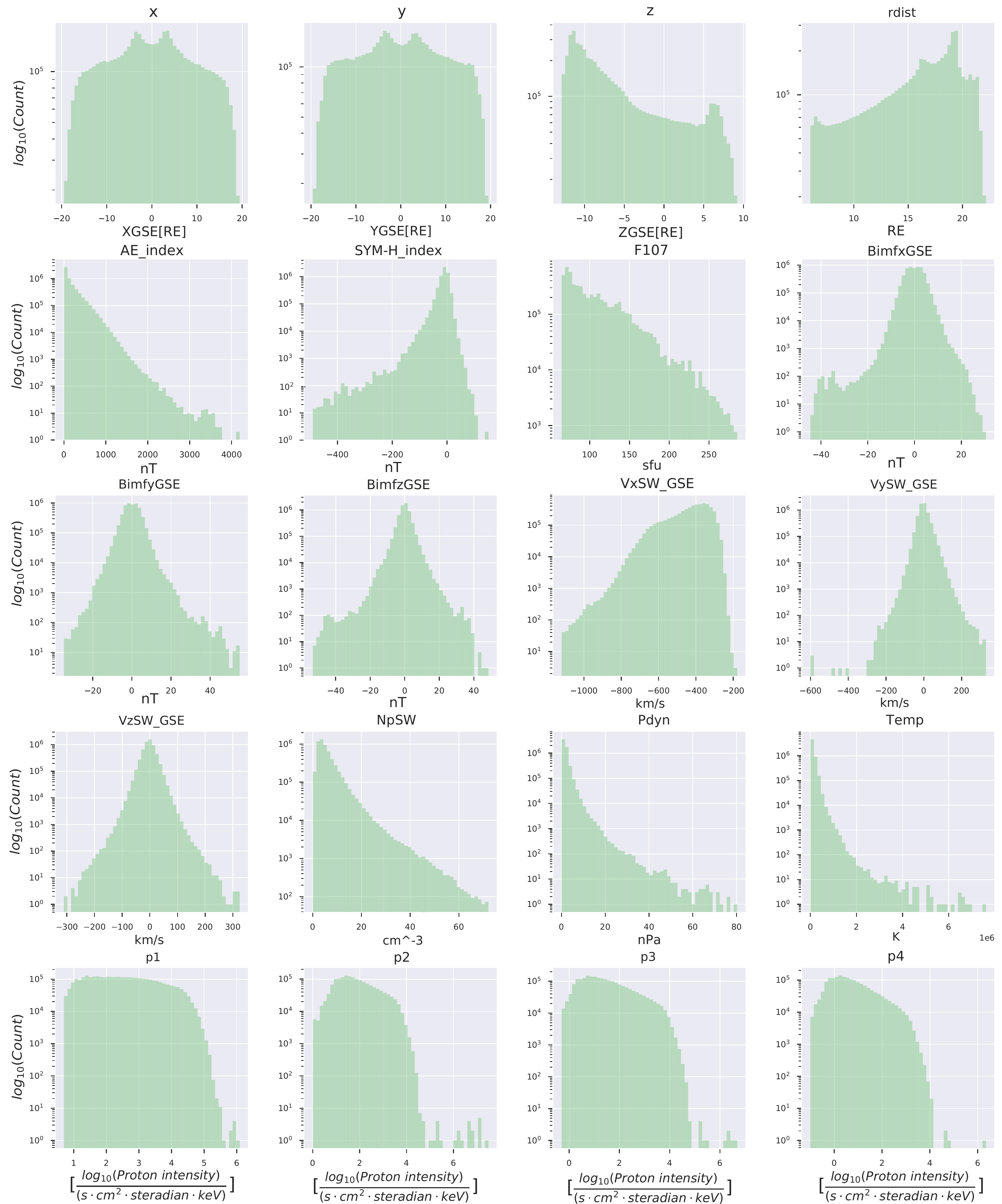}
 	\caption{Histogram of the number of samples of predictors and proton intensities used in the model. The intensities per channel are in $log_{10}$ space. \label{fig:hist} }
 \end{figure}

\appendix
Figure~\ref{fig:hist} shows the distribution of the number of samples for the predictors and the proton intensities (on the vertical axis) with a given value range (on the horizontal axis).

Table~\ref{tab:hpo} shows the search space for hyper-parameter optimization and the best values found for the MLP model.
\begin{table}
    \centering
    \begin{tabular}{lll}
    \toprule
    Parameter & Search Space & Best Value \\
    \midrule
    \multicolumn{3}{c}{Multilayer Perceptron} \\
    \midrule
    Number of Layers & [2, 5] & 2 \\
    Hidden Dimensions & [32, 256] & [64, 32] \\
    \midrule
    \multicolumn{3}{c}{Optimizer} \\
    \midrule
    Batch Size & [32, 256] & 56 \\
    Learning Rate & [$10^{-6}$, $10^{-3}$] & 0.00046 \\
    L2 Penalty & [0, $10^{-2}$] & 0.00882 \\
    \bottomrule
    \end{tabular}
    \caption{Search space for hyper-parameter optimization and the best values in the MLP model.}
    \label{tab:hpo}
\end{table}

\end{document}